# CT dose reduction factors in the thousands using X-ray phase contrast

Marcus J. Kitchen[1]*, Genevieve A. Buckley[1], Timur E. Gureyev[2,3,1], Megan J. Wallace[4,5], Nico Andres-Thio[6,7], Kentaro Uesugi[8], Naoto Yagi[8] and Stuart B. Hooper[4,5].

[1]School of Physics and Astronomy, Monash University, Melbourne, Australia.

[2]ARC Centre of Excellence in Advanced Molecular Imaging, School of Physics, University of Melbourne, Victoria, Australia.

[3]School of Science and Technology, University of New England, Armidale, NSW 2351, Australia.

[4]The Ritchie Centre, Hudson Institute for Medical Research, Melbourne, Australia.

[5]Department of Obstetrics and Gynaecology, Monash University, Melbourne, Australia.

[6]School of Engineering, Melbourne University, Victoria, Australia.

[7] School of Mathematics and Statistics, Melbourne University, Melbourne, Australia.

[8]Japan Synchrotron Radiation Research Institute (JASRI/SPring-8), 1-1-1 Kouto, Sayo, Hyogo 679-5198, Japan.

* All correspondence should be addressed to Marcus.Kitchen@monash.edu

**Phase-contrast X-ray imaging can improve the visibility of weakly absorbing objects (e.g. soft tissues) by an order of magnitude or more compared to conventional radiographs. Previously, it has been shown that combining phase retrieval with computed tomography (CT) can increase the signal-to-noise ratio (SNR) by up to two orders of magnitude over conventional CT at the same radiation dose, without loss of image quality. Our experiments reveal that as radiation dose decreases, the relative improvement in SNR increases. We discovered this enhancement can be traded for a reduction in dose greater than the square of the gain in SNR. Upon reducing the dose 300 fold, the phase-retrieved SNR was still almost 10 times larger than the absorption contrast data. This reveals the potential for dose reduction factors in the tens of thousands without loss in image quality, which would have a profound impact on medical and industrial imaging applications.**

X-ray radiography and computed tomography (CT) are two of the most common imaging modalities in diagnostic medicine. However, soft tissues have similar X-ray absorption properties resulting in poor contrast obscured by noise. Photon (Poisson) noise can be reduced by increasing radiation dose, but dose must be minimised for patient safety. New techniques have been developed to enhance contrast by an order of magnitude or more[1-3] using phase shifts (i.e. refraction) of X-rays.

Propagation-based imaging (PBI) is the simplest phase contrast technique, relying on Fresnel diffraction[4], and well-suited to polychromatic, micro-focus X-ray sources[5]. Fresnel diffraction simultaneously enhances image contrast and spatial resolution[6] with negligible increase in image noise.

Recovering an object's complex refractive index is difficult since only intensity, not phase, is measured directly. Most phase retrieval algorithms place restrictions on the object's composition and/or are unstable against image noise[7]. Paganin et al.[8] developed an especially noise-robust algorithm for homogenous, single-material samples (TIE-Hom; see Methods) [9-11]. Beltran et al.[12,13] extended TIE-Hom for multi-material samples. These algorithms have been shown to improve the signal-to-noise ratio (SNR) by up to 200 fold over conventional CT with minimal loss of spatial resolution[12-14].

If image noise contains photon noise only, the SNR is proportional to the root of the mean photon number per pixel, $\bar{n}$ [15]. Since photon number is proportional to radiation dose and, in the case of constant photon rate, $\eta$, also to exposure time, $t$, as $\bar{n} = \eta t$, the gain in SNR is:

$$G(z) = \frac{SNR_{PR}}{SNR_{AC}} = \frac{\mu}{\left[\sqrt{\eta t}/G_P(z)\right]} \times \frac{\sqrt{\eta t}}{\mu} = \frac{\sigma_{AC}}{\sigma_{PR}} = G_P(z) \qquad (1)$$

The attenuation coefficient, $\mu$, is the CT signal. PR and AC denote Phase Retrieved and Absorption Contrast data. The standard deviation of Poisson noise, $\sigma_P = \sqrt{\eta t}$, is normalised during flat field correction, hence noise in the corrected images is inversely proportional to the mean photon number per pixel: $\sigma_{P,corr} = \sqrt{\eta t}/(\eta_0 t) = 1/\sqrt{\eta' t}$, where $\eta_0$ is the photon rate in the incident beam and $\eta' = \eta_0^2/\eta$. Normalization doesn't change equation (1), as the same factors, $\bar{n}_0 = \eta_0 t$, in the numerator and denominator cancel out. $G_P(z)$ is the factor by which Poisson noise is reduced by phase retrieval. This gain in SNR can be traded for a reduced radiation dose. The dose reduction factor (DRF) is calculated by equating $G(z)$ to the ratio of SNRs for phase retrieval at the original exposure time $t$ to that of a shorter time, $t'$, assuming constant dose rate, so $SNR_{PR}(t') = SNR_{AC}(t)$. Considering photon noise only, equation (1) gives a DRF of:

$$DRF = \frac{Dose_1}{Dose_2} = \frac{t}{t'} = \frac{SNR_{PR}^2(t)}{SNR_{PR}^2(t')} = \frac{SNR_{PR}^2(t)}{SNR_{AC}^2(t)} = G_P^2(z). \tag{2}$$

Nesterets *et al.*[16] showed the gain factor $G_P(z)$ is larger for CT than for a projection image, hence the potential for dose reduction is much greater for CT than for radiography. Equation (2) suggests that with SNR gains in the hundreds being possible, a DRF in the tens of thousands is also possible. However, since other noise sources are typically present (e.g. detector noise and reconstruction artefacts[14]), this analysis is too simple. This study determines how low the dose can be in propagation-based CT without losing image quality.

## Results

A newborn rabbit thorax was imaged for this study to show the impact of phase retrieval on improving the image SNR whilst maintaining sufficiently high spatial resolution to resolve individual alveoli (the smallest lung airspaces).

Figure 1 shows a reconstructed slice through the lungs at the smallest and largest propagation distances, and the effect of applying TIE-Hom phase retrieval algorithm before CT reconstruction. Figure 2 shows close-up images of the terminal airways. At the larger distance L = 2 m (Figure 1b), phase contrast provides high contrast halos (fringes) at each air/tissue and bone/tissue interface. This edge enhancement increases the spatial resolution[17] with neglible increase in noise (Extended Data Figure 1) and the high fringe contrast is well above the noise floor. The phase retrieved image (Figure 1c) shows that both the phase contrast halos and high frequency noise have been suppressed by TIE-Hom (Extended Data Figure 2). The gain in spatial resolution from Fresnel diffraction has been traded for an increase in the SNR.

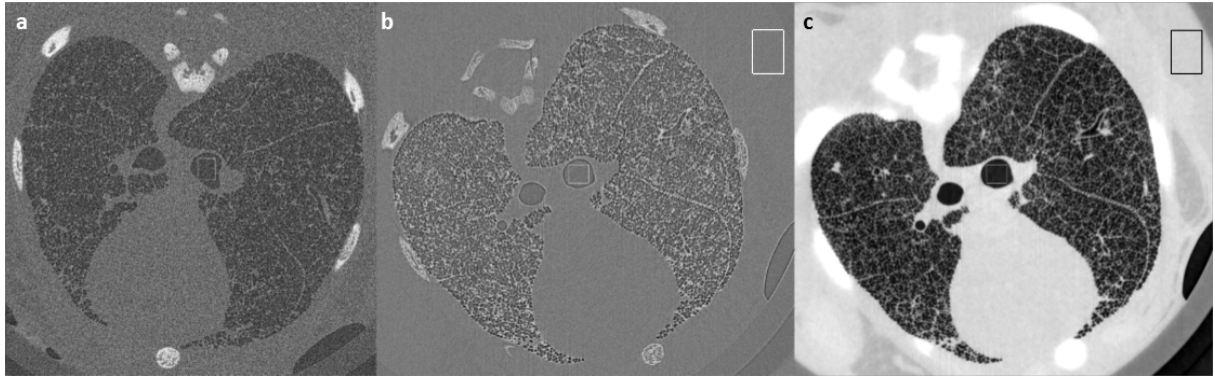

**Figure 1: CT slice reconstruction of rabbit kitten lungs. a**, Absorption contrast CT reconstruction at sample to detector distance 0.16 m. **b**, 2 m; **c**, and with phase retrieval (TIE-Hom) at 2 m. The dark areas represent air-filled airways in the lungs and bones appear bright. Black and white boxes indicate typical positions of uniform regions of interest for SNR analysis. The exposure time was 10 ms per projection for all images. Separate greyscale palettes have been used for each image. Image dimensions: **a**, 18.36 mm × 18.51 mm; **b** and **c**, 20.65 mm x 18.51 mm.

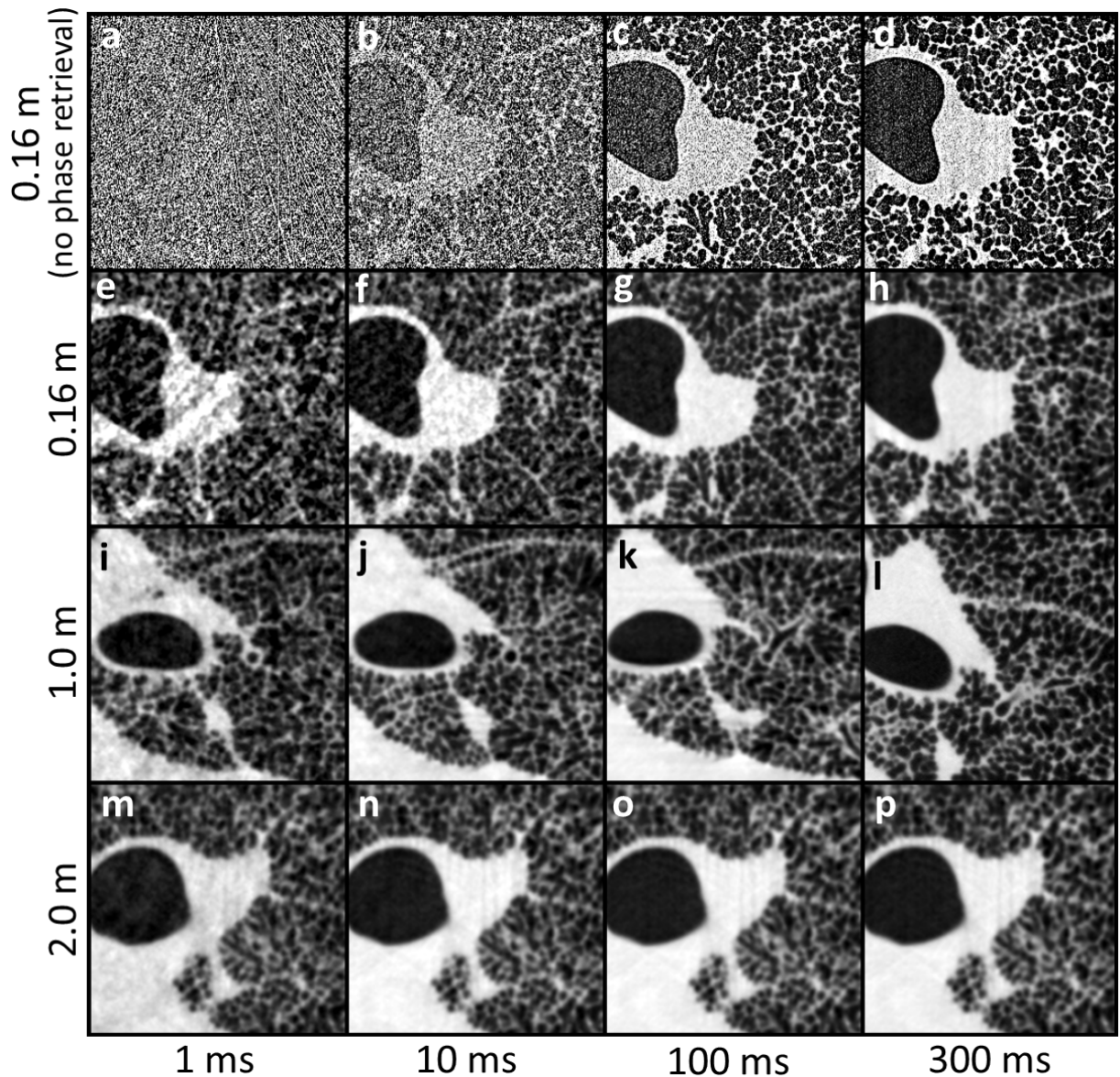

**Figure 2: Magnified lung tissue reconstructions as a function of propagation distance and exposure time (time stated per projection).** The same greyscale palette has been used for all images. Image dimensions: 3.83 mm × 3.83 mm. Dark regions are cross-sections through the airways including large bronchioles and alveoli (~160 μm diameter)[18]. TIE-Hom retrieval has been employed for all but the top row of data. An increase in image quality of soft tissues is observed as both variables (propagation distance, and exposure time) increase. At the shortest distance (0.16 m) the raw reconstructions (no phase retrieval) are dominated by noise at the two shortest exposure times, but at the two longest exposure times individual alveoli are clearly visible. Even at 0.16 m, phase contrast halos highlight the airways and enhance their apparent spatial resolution, showing that this is not a true

absorption contrast image. A substantial improvement in image quality is also seen when phase retrieval is applied, even at 0.16 m. Phase retrieval removes the halo artefacts and greatly suppresses noise without losing visibility of the microscopic alveoli, even for the 1 ms exposures. At larger distances the image quality appears remarkably consistent across all exposure settings, despite the dose varying by a factor of 300.

**Signal-to-noise ratio measurements**

Image quality was measured quantitatively using the signal-to-noise ratio (SNR) from small, homogeneous regions of interest in the agarose surrounding the animal (Figure 1). Agarose has similar X-ray absorption and refraction properties to the biological tissue it surrounds and provides a homogeneous medium that is necessary for the quantitative analysis of noise. These results are presented in Table 1.

**Table 1: SNR from agarose regions of interest.** Agarose SNR data with and without phase retrieval. Uncertainties are the standard deviation of measurements from five neighbouring CT slices.

| Exposure time (ms) | Phase retrieved data Sample-to-detector distance (m) | | | No phase retrieval Sample-to-detector distance (m) | | |
|---|---|---|---|---|---|---|
| | 0.16 | 1.0 | 2.0 | 0.16 | 1.0 | 2.0 |
| 1 | 3.77 ± 0.03 | 15.23 ± 0.03 | 28.3 ± 0.3 | 0.15 ± 0.01 | 0.14 ± 0.02 | 0.127 ± 0.005 |
| 10 | 9.99 ± 0.08 | 39.3 ± 0.5 | 61.0 ± 0.4 | 0.575 ± 0.002 | 0.537 ± 0.006 | 0.507 ± 0.009 |
| 100 | 30.5 ± 0.6 | 86.7 ± 0.6 | 93.0 ± 0.8 | 1.758 ± 0.007 | 1.41 ± 0.02 | 1.25 ± 0.01 |
| 300 | 49.6 ± 0.8 | 97.4 ± 0.6 | 103 ± 1 | 2.95 ± 0.02 | 1.993 ± 0.008 | 1.59 ± 0.03 |

Increasing the exposure time improved the SNR approximately proportional to the square root of time, as expected from Poisson statistics. Increasing the propagation distance and using phase retrieval also improved the SNR. We next look at the gain in SNR to determine how far the exposure time, hence radiation dose, can be reduced using phase retrieval.

**Measurements and analysis of SNR gain**

Figure 3a shows the gain in SNR for phase retrieved data in agarose as a function of propagation distance, at each exposure time, with respect to the absorption contrast data at the same exposure time. Figure 3b shows the same data plotted against exposure time. We discover an unexpected effect that the SNR gain is consistently highest at the shortest exposure times. Achieving gains in SNR in the hundreds at short exposure times has great potential for high throughput applications or low dose biomedical applications.

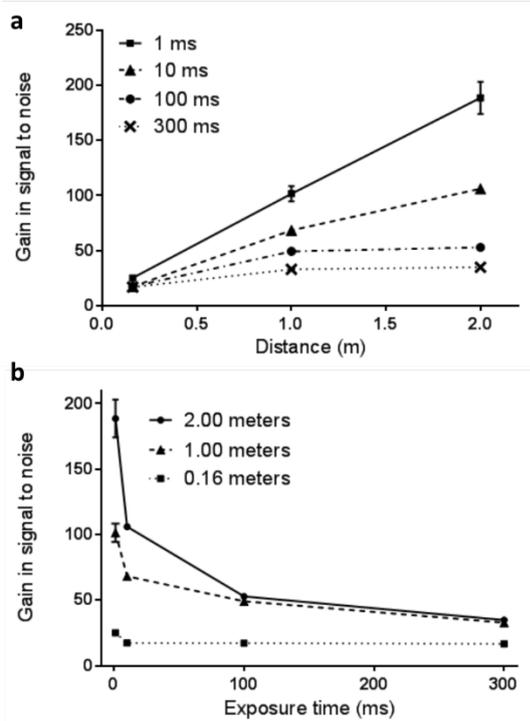

**Figure 3: Plots of gain in SNR; phase retrieved data with respect to absorption contrast data, calculated separately at each exposure time. a,** Gain in SNR vs propagation distance. Exposure times per projection: 1 ms (squares, solid line), 10 ms (triangles, dashed line), 100 ms (circles, dash-dotted line), and 300 ms (crosses, dotted line). **b,** Gain in SNR vs exposure. Sample to detector distances were 0.16 m (squares, dotted line), 1.0 m (triangles, dashed line), and 2.0 m (circles, solid line). Gain uncertainties were calculated by summing fractional SNR uncertainties from phase- and absorption-contrast data, and scaled by the gain value.

**Mathematical basis for SNR gain**

Previously, Gureyev et al.[14] proposed that the gain in SNR in propagation-based CT with TIE-Hom phase retrieval can reduce because of CT reconstruction artefacts (e.g. ring or streak artefacts), which modify equation (1) to:

$$G(z) = \sqrt{\frac{\sigma_{P,corr}^2 + \sigma_{CT\_art}^2}{[\sigma_{P,corr}^2/G_P^2(z)] + \sigma_{CT\_art}^2}} = G_P(z)\sqrt{\frac{1+\kappa^2}{1+G_P^2(z)\kappa^2}} \quad (3)$$

where $\kappa = \sigma_{CT\_art}/\sigma_{P,corr}$. Noise from CT artefacts is $\sigma_{CT\_art}$, and Poisson noise from normalized projection images is $\sigma_{P,corr}$. Equation (3) assumes that CT artefacts are identical for phase retrieved and absorption contrast data. At large propagation distances the TIE may be invalid and introduce different artefacts for the phase retrieved data, however TIE-Hom can also reduce detector noise and some CT reconstruction artefacts, due to its low-pass filtering effect. Equation (3) also does not account for other sources of noise such as detector dark current and read noise that may be present. In equation (4) below we extend equation (3) to account for these effects. Since our sCMOS detector (see Methods) showed no time-dependent dark current noise we consider only time-independent noise. Assuming noise sources to be statistically independent allows us to add the corresponding variances, equation (3) then becomes:

$$G(z,t) = \sqrt{\frac{\sigma_{P,corr}^2 + \sigma_D^2 + \sigma_{AC\_art}^2}{[\sigma_{P,corr}^2/G_P^2(z)] + (\sigma_D^2/G_D^2) + \sigma_{PR\_art}^2}} \quad (4)$$

$$= G_P(z)\sqrt{\frac{1+\kappa_{AC}^2(z=0,t)}{1+G_P^2(z)\kappa_{PR}^2(z,t)}}$$

Here subscript 'P'= Poisson, 'D'= Detector and 'art'= CT artefacts; $\sigma_D$ is the time-independent detector noise; $\kappa_{AC}^2 \equiv (\sigma_D^2 + \sigma_{AC\_art}^2)/\sigma_{P,corr}^2$; and $\kappa_{PR}^2 \equiv [(\sigma_D^2/G_D^2) + \sigma_{PR\_art}^2]/\sigma_{P,corr}^2$. $G_D$ represents suppression of detector noise by TIE-Hom. This is treated separately from Poisson noise as the spatial frequency distribution of each will be different.

The data seen in Figure 3 can all be explained using qualitative analysis of equation (4). When Poisson noise dominates at the shortest exposure time and the Poisson gain factor $G_P$ is not large (i.e. $\kappa_{AC}^2 \ll 1$ and $\kappa_{PR}^2 G_P^2 \ll 1$), equation (4) shows that $G(z,t) \approx G_P(z)$. Nesterets et al.[16] showed that $G_P(z) \cong \gamma \lambda z / h^2$, where $\gamma = \delta/\beta$ is the ratio of the real to imaginary parts of the decrement of the (relative) X-ray refractive index, $\lambda$ is the X-ray wavelength and $h$ is typically 0.5 times the full width half maximum (FWHM) of the detector point spread function. This explains the linear relationship with distance at the shortest exposure time in Figure 3a. Conversely, when the CT artefacts and detector noise dominate the Poisson noise (i.e. for large exposure times when $\kappa_{AC}^2, \kappa_{PR}^2 \gg 1$) then $G(z,t) \approx \kappa_{AC}/\kappa_{PR}$, which is independent of time. This explains the relatively flat lower curve in Figure 3a at the longest exposure time and why the curves in Figure 3b converge at large exposure times. Finally, when Poisson noise is dominant, but $G_P(z)$ is large (i.e. $\kappa_{AC}^2 \ll 1$ and $\kappa_{PR}^2 G_P^2 \gg 1$) then $G(z,t) \approx 1/\kappa_{PR} \cong 1/\sqrt{\eta' t [(\sigma_D^2/G_D^2) + \sigma_{PR\_art}^2]}$. This explains the inverse relationship between SNR gain and exposure time seen in Figure 3b.

With this new model for gain in SNR (equation (4)), the dose reduction factor can be calculated according to equation (2). Substituting $1/t' = DRF/t$ into the equality $SNR_{AC}(t) = SNR_{PR}(t')$ and solving the resultant equation for $DRF$, we obtain:

$$DRF(z,t) = G_P^2(z)\{1 + \eta' t [\sigma_{AC_{art}}^2 - \sigma_{PR_{art}}^2 + \sigma_D^2(1 - 1/G_D^2)]\} \qquad (5)$$

Equation (5) shows that, if $\sigma_{AC_{art}}^2 > \sigma_{PR_{art}}^2$ and $G_D^2 > 1$, as would normally be the case, then $DRF > G_P^2$. This is an unexpected result, showing that the dose reduction factor due to PBI CT with TIE-Hom retrieval in the presence of detector and reconstruction artefacts can be higher than in the ideal case with only Poisson noise present. This is explained by the fact that the TIE-Hom retrieval not only reduces the Poisson noise, but also suppresses the detector noise and some CT reconstruction artefacts via the same low-pass filtering effect. On the other hand, in some cases TIE-Hom phase retrieval may introduce new artefacts, e.g. due to strong deviations of the sample from the homogeneous model (even though such artefacts are expected to be weak in most situations[19]). In such cases it is possible that $\sigma^2_{AC_{art}} < \sigma^2_{PR_{art}}$ and, if the detector noise is weak, then the *DRF* can be actually smaller than $G_P^2$. In most practical cases, though, it should be expected that $DRF \approx G_P^2$, even when the gain factor $G$ is significantly smaller than $G_P$. As a consequence, $DRF > G^2$ in particular. These equations show that the dose reduction can indeed be in the tens of thousands if the propagation distance is sufficiently large and the beam energy is sufficiently low.

Equation (5) also shows that the DRF increases with exposure time $t$. This is because $G(z,t) \propto 1/\sqrt{t}$, so at short exposures a small reduction in time gives a large increase in SNR whilst at long exposures a larger time reduction is required to get the same increase in SNR.

**Experimental dose reduction factors**

We can estimate the DRFs in two different ways from the experimental data.

*Method 1:* When the CT acquisition and reconstruction artefacts are small compared to Poisson noise (i.e., at low radiation dose), $DRF \approx G_P^2(z)$. Figure 3a shows that the 1 ms data most closely approximates the Poisson limit for which the other parameters are negligible, hence $G(z,t) \approx G_P(z)$. At 1 ms exposures $G_P(z) = 25 \pm 2$; $101 \pm 7$; and $189 \pm 15$ for $z = 0.16$ m, 1.0 m and 2.0 m, respectively. Thus the DRFs are thus approximately $G_P^2(z) = 625$; 10,200; and 35,700.

*Method 2:* At all propagation distances, the dose was reduced by a maximum of 300 fold in the experiment due to detector limitations (see Methods). Even after this reduction the phase retrieved images had a higher SNR that the absorption contrast data. The SNR of the 1 ms phase retrieved data was still larger than the absorption contrast SNR at 300 ms by factors of $1.28 \pm 0.02$ (= 3.77/2.95); $5.16 \pm 0.05$ (= 15.23/2.95); and $9.6 \pm 0.2$ (= 28.3/2.95) at 0.16 m, 1.0 m and 2.0 m, respectively (see Table 1). From equation (5) we can estimate the remaining dose reduction factor as the square of these numbers. This gives the expected dose reduction factors at 0.16 m, 1.0 m and 2.0 m of $300 \times 1.28^2 = 490 \pm 20$; $300 \times 5.16^2 = 7,990 \pm 30$; and $300 \times 9.6^2 = 27,600 \pm 30$, respectively.

We see that both methods give similar estimates for the DRF at each distance, with the first method providing an upper bound and the second method giving more accurate figures according to the nature of approximations used in the two methods.

## Discussion

Herein we have shown experimentally that dose reduction factors much greater than the 300 fold reduction we employed experimentally are possible using PBI CT combined with a collimated X-ray beam and the phase retrieval algorithm of Paganin *et al.*[8] (TIE-Hom). Using the derivation of Nesterets *et al.*[16], the maximum limit to the SNR gain in the absence of detector and CT reconstruction artefacts is $0.3\gamma$ (see Methods), which equals 532 for our experiment. Our model for noise analysis shows that this optimal condition could lead to a dose reduction factor of up to $\sim(0.3\gamma)^2 = 532^2 = 283,024$ fold. This remains to be verified experimentally and will likely require high-efficiency photon-counting detectors with a large dynamic range.

We unexpectedly discovered that the gain in SNR with TIE-Hom is significantly larger at low radiation doses when Poisson noise dominates the images. This phenomenon can be explained by accounting for all sources of noise in the reconstruction (equation (4)). These findings will have important consequences for CT studies, particularly for pre-clinical studies whereby large dose reductions can be achieved for longitudinal *in vivo* studies.

With dose reduction factors larger than the number of projections (here 1,800), we find that, using phase-contrast imaging and phase retrieval, CT with high SNR and spatial resolution can potentially be achieved with less dose than a single projection absorption-based image. We therefore recommend using a large number of very low dose projections, coupled with phase retrieval before CT slice reconstruction. This will result in images with high SNR, retaining high spatial resolution, and minimizing any reconstruction artefacts due to insufficient CT projection angles.

We are currently translating these findings to more readily available laboratory micro-focus sources for widespread use. The challenge for such sources, apart from the lower brightness compared to synchrotrons, is that geometric magnification of the divergent point source shortens the effective propagation distance, thereby reducing phase contrast[11]. We nevertheless anticipate dose reduction factors in the hundreds as we found for our shortest propagation distance of 16 cm.

The ability to improve CT image quality by factors in the tens to hundreds, or to reduce radiation exposure by factors in the hundreds to thousands, would have a dramatic impact in both commercial and diagnostic imaging applications. Using less radiation will enable higher throughput imaging with fewer motion artifacts and be safer for human imaging or for longitudinal preclinical studies. Diagnostic imaging of the lungs would likely benefit significnatly since the air tissue boundaries produce the greatest phase contrast within the mammalian body[20]. This technique also has great potential for mass breast cancer screening using tomography with high resolution and with very low radiation dose[19]. Finally, the demonstrated dose reduction also lowers the requirements for brightness of micro-focus X-ray sources that can be used for medical phase-contrast X-ray imaging, thus potentially opening the way for the introduction of this method into routine clinical practice.

## Methods

**Phase retrieval**

For X-ray interactions, objects can be defined in terms of the wavelength-dependent three-dimensional (3D) complex refractive index: $n(\mathbf{r}, \lambda) = 1 - \delta(\mathbf{r}, \lambda) + i\beta(\mathbf{r}, \lambda)$, where **r** is a 3D position vector and $\lambda$ is the wavelength. The real decrement dictates phase changes (refraction) within the object and the imaginary term describes attenuation of the beam. The linear attenuation coefficient $\mu(\lambda) = 2k\beta(\lambda)$, where $k = 2\pi/\lambda$. Paganin et al.[8] showed a quantitative reconstruction of an object can be achieved from a single propagation-based phase contrast image of a "homogenous"[8] or "monomorphous"[21] object using the Transport-of-Intensity equation (TIE)[22]. For such an object the ratio of $\delta(\mathbf{r}, \lambda)/\beta(\mathbf{r}, \lambda) \equiv \gamma$ must be constant throughout the material. Note that, at energies far from absorption edges, $\delta$ and $\beta$ are proportional to electron density, hence $\gamma$ will be independent of density. The TIE is valid only for relatively small sample-to-detector propagation distances, such that no more than a single Fresnel fringe pair should be seen at the boundaries between objects where the phase gradients are strongest. Under these conditions, and in the case of plane wave illumination, the intensity map at a distance $z = L$ beyond the object plane ($z = 0$) is[8]:

$$I(\mathbf{r}_\perp, z = L) = \left(1 - \frac{L\gamma}{2k}\nabla_\perp^2\right)I(\mathbf{r}_\perp, z = 0). \tag{6}$$

Here I is the intensity, $\mathbf{r}_\perp$ is the position vector in the plane perpendicular to the optic axis $z$, and $\nabla_\perp^2$ denotes the Laplacian operator in that plane. The Laplacian operator amplifies high spatial frequencies (intensity gradients) in the image caused by refraction at the object boundaries. One may expect that Poisson noise in the image will also be amplified by this Fresnel diffraction[16]. However, experimentally we find that the noise changes negligibly upon propagation in the parallel beam geometry when intensity is conserved, as demonstrated in

Extended Data Figure 1. This is the key gain that phase contrast provides, but the reason behind this conservation of noise is still under investigation. The conventional explanation of this phenomenon uses the fact that image noise appears only in the process of photon detection, and it depends primarily on the statistics of the photon fluence and the detector properties. If the same detector is used for image registration in the "contact" ($z=0$) and "propagated" ($z=L$) planes, the X-ray absorption in air between the two planes is negligible (otherwise an evacuated beam pipe can be used), and the image contrast is relatively weak (as required by the validity conditions of the TIE-Hom), then the detection conditions are generally equivalent in the two planes. Therefore, in this case the noise should be the same in both planes, as observed in the experiments.

Given an intensity map at plane $z = L$, the contact plane ($z = 0$) intensity map can be recovered as[8] (TIE-Hom retrieval):

$$I(\mathbf{r}_\perp, z = 0) = \mathbf{F}^{-1}\left\{\frac{1}{(\pi\gamma\lambda L \mathbf{k}_\perp^2) + 1} \mathbf{F}[I(\mathbf{r}_\perp, z = L)]\right\}. \tag{7}$$

Here $\mathbf{F}$ and $\mathbf{F}^{-1}$ represent Fourier and inverse Fourier transform operators, respectively, $\mathbf{k}_\perp^2$ is the squared Fourier space radius dual to $\mathbf{r}_\perp$. A similar formalism has been derived for point source illumination[8,23]. Equation (7) is a low-pass Fourier filter that smooths out the contrast at high spatial frequencies that was amplified by Fresnel diffraction. This filter is highly robust against noise[7,10] since the denominator is never zero, even when $\mathbf{k}_\perp^2 = 0$. Extended Data Figure 2 shows the effect of signal and noise suppression in Fourier space.

Recovery of the 3D complex refractive index of the object requires a tomographic projection series to be recorded with the sample rotated around some axis in the plane $(\mathbf{r}_\perp, z = 0)$, and the detector fixed at plane $z = L$. TIE-Hom retrieval must be applied to each projection image before using any choice of tomographic reconstruction, such as filtered back-projection[24]. This process can very accurately reconstruct sample's $\beta$ values, but in certain cases will give non-quantitative reconstructions of the $\delta$ value[19].

Beltran *et al.*[12] extended the work of Paganin *et al.*[8] to enable objects comprised of multiple materials to be reconstructed one pair of materials at a time. They demonstrated that when the choice of constants for the material pairing is correct, those materials will be correctly reconstructed. However, other materials in the samples will be locally polluted by incorrect filtration resulting in under or over-smoothing of the material boundaries[12,21]. Those studies show that an arbitrary choice of low pass filter will not correctly suppress the phase contrast, nor preserve the spatial resolution. Hence one cannot simply employ typical image filtering routines to quantitatively recover the object, particularly when using a multi-material sample.

Nesterets *et al.*[16] showed that under the approximation of the TIE, the maximum improvement in SNR by using phase contrast in conjunction with TIE-Hom for CT is $\sim 0.3\gamma$. This variable is typically in the hundreds for low density materials in the diagnostic energy range, They also showed that the gain in SNR with TIE-Hom is greater for CT data than for individual projections. Equation (2) suggests that the potential for dose reduction should thus be higher for CT than for a 2D projection imaging.

While in theory, when the "near-Fresnel" imaging conditions are satisfied (which is the case in the present experiment), the TIE-Hom retrieval should restore the spatial resolution achievable in the contact plane under the equivalent imaging conditions (i.e. using the same X-ray illumination and same detector)[8,11]. In practice, however, there is often a moderate loss of resolution associated with this method (Extended Data Figure 3). This effect is mostly due to the deviation of the sample composition from the homogeneous (single-material) one assumed in the TIE-Hom method, which can lead to local blurring of the edges and interfaces in the sample. This loss of spatial resolution can be minimised by careful selection of the experimental conditions (the propagation distance, the source size and the detector resolution) and by judicious choice of the parameter $\gamma$ in TIE-Hom retrieval.

Spatial resolution of the CT images (Extended Data Figure 3) was calculated from the line profiles of the sharp edge of the PMMA tube in air. This was represented by the convolution of a Heaviside step function with a Gaussian defined by its full width half maximum (FWHM). Note that here we assumed the material was water for providing a good approximation to tissue, yet the resolution was measured from the plastic tube, hence the $\gamma$ value was not ideal for use in that location.

**Animal handling**

This experiment used a newborn New Zealand White rabbit kitten that had been used in experiments conducted with approval from the SPring-8 Animal Care (Japan) and Monash University (Australia) Animal Ethics Committees. The kitten was humanely killed in line with approved guidelines and the carcass scavenged for this experiment. The lungs of the kitten were inflated *in situ* with nitrogen using a sustained volume of 20 ml/kg before ligating the trachea. Nitrogen was used to prevent post-mortem changes in the lung air gas volume as oxygen can diffuse into the surrounding tissues. The body of the rabbit kitten was then set in a 2% agarose solution to prevent movement during multiple CT acquisitions.

**Synchrotron experimental imaging**

Experimental data was acquired in hutch 3 of beamline 20B2 at the SPring-8 synchrotron Japan, 210 m from the 150 μm(H)×10 μm(V) source. A beam energy of 24 keV was selected using a double bounce Si(111) monochromator. A Hamamatsu ORCA flash C11440-22C sCMOS detector was coupled to a Nikkon 85mm lens and a 25 µm thick Gadox phosphor (*P43, $Gd_2O_2S:Tb^+$*). The detector field of view was 2048 x 2048 pixels (31.3 x 31.3 mm) and the effective pixel size was 15.3 µm. The detector dark noise was negligible at ~0.15 electrons/pixel/second. The readout noise was also low at 1.3 electrons per exposure (Hamamatsu C11440-22C instruction manual). The dose rate (air kerma) to the sample was kept fixed at 13.5±0.1 mGy/s, as measured using an air-filled ionization chamber.

Twelve separate computed tomography (CT) datasets of the same animal thorax were acquired to investigate the dependence of image quality on: (1) the sample-to-detector propagation distance ($L$); (2) the exposure time ($t$), and; (3) the effect of applying TIE-Hom retrieval to the projection images before CT reconstruction. CT datasets were acquired at propagation distances of 0.16 m, 1.0 m and 2.0 m, and using four different exposure times of 1 ms, 10 ms, 100 ms, and 300 ms per projection at each distance. The shortest distance of 0.16 m was the closest we could safely position the detector to minimise phase contrast. This exposure range was chosen since the shortest exposure time (1ms) was the lower limit of the detector using external triggering, and exposures longer than 300 ms saturated the detector. A total of 1801 projections was recorded per scan as the sample was rotated through 180°. This gave a surface entry radiation dose range from 24.3±0.1 mGy (1 ms exposures) to 7.29±0.01 Gy (300 ms exposures). We note that the lower dose is comparable to clinical CT scanners[15], but with much higher spatial resolution, whilst the largest dose is well above safe limits for clinical imaging but is comparable to typical micro-CT scans on non-living samples[25-27].

**Image processing for CT reconstruction**

After the acquisition of experimental data, all images were corrected for the detector dark current offset and sample images were normalised by the flat field intensity (beam with no sample). Two separate image pre-processing cases were investigated: filtered back-projection[24] (FBP; using a ramp filter) reconstruction with no further image pre-processing, and FBP reconstruction following the application of TIE-Hom to each of the projection images. Here we assume that all tissue types contained within the thorax have a $\gamma$-value equivalent to water

($\beta = 2.25 \times 10^{-10}$; $\delta = 3.99 \times 10^{-7}$; given 24 keV X-rays). This is a reasonable approximation since for thoracic imaging we are primarily concerned with the contrast between lung air gas and the surrounding tissue, and not subtle differences between tissue types. It is also well-known that the TIE-Hom reconstruction is quite insensitive to variations of the value of parameter $\gamma$. Using the approximation of Nesterets et al.[16], the maximum limit to the SNR gain due to propagation-based phase contrast and TIE-Hom retrieval for these parameters is $0.3\gamma = 532$.

**Determining unknown parameters in DRF**

To determine the unknown parameters in equation (4), numerical curve fitting techniques were applied to the experimental SNR data. However, the best fits based on our model for SNR could never consistently fit all of the points, typically underestimating the SNR at lowest exposures. Since spatial resolution can significantly affect the SNR, we measured the resolution of the phase retrieved images for all distances and exposure times. For this beamline image blurring from the finite source size should be negligible at all used distances ($< 0.1 \times$ pixel size). We made the surprising discovery that the spatial resolution varied with both propagation distance and time.

Extended Data Figure 3 shows the spatial resolution to degrade (larger FWHM) at longer propagation distances (this is discussed above in the "Phase retrieval" section). More importantly, we see the resolution improve (reduced FWHM) for shorter exposure times. This is possibly due to an increase in the effective source size resulting from motion of the beam, likely caused by monochromator vibration, during the exposure. However, in principle, blurring from source should only affect edge sharpness and should not affect the SNR, as evidenced by Extended Data Figure 1. We speculate then that parameterisation problem comes from either insufficient data points, variability in the detector noise with each readout, or inaccuracy of the detector timing control at short exposure times. In future we will reduce the radiation dose by decreasing the photon flux rather than altering exposure times to rule out these problems.

**Extended Data** is linked to the online version of the paper.

# Acknowledgements

We thank David M. Paganin for discussions when developing this manuscript. M.J.K. is supported by an ARC Future Fellowship (FT160100454). S.B.H. is an NHMRC Principal Research Fellow. This work was funded by ARC Discovery Projects DP110101941 and

DP130104913, and supported by SPring-8 proposals 2014B1522 and 2014A0047. Animal ethics number MMCA/2014/15 (Monash Medical Centre Committee A).

## Author contributions

M.J.K., G.A.B., M.J.W. and S.B.H. conducted the experiment with assistance and expertise from K.U. and N.Y.; analysis of data by M.J.K., G.A.B. and N.A.-T.; theoretical contributions by M.J.K. and T.E.G.; and the manuscript was written by M.J.K., G.A.B. and T.E.G.

## Author information

Data analyzed in this paper have been deposited into the Store.Monash repository, identifiable by the doi: 10.4225/03/58197dd586bef. Reprints and permissions information is available at [www.nature.com/reprints](www.nature.com/reprints). The authors have no competing financial interests to declare. Correspondence and requests for materials should be addressed to [Marcus.Kitchen@monash.edu](Marcus.Kitchen@monash.edu)

**Extended Data**

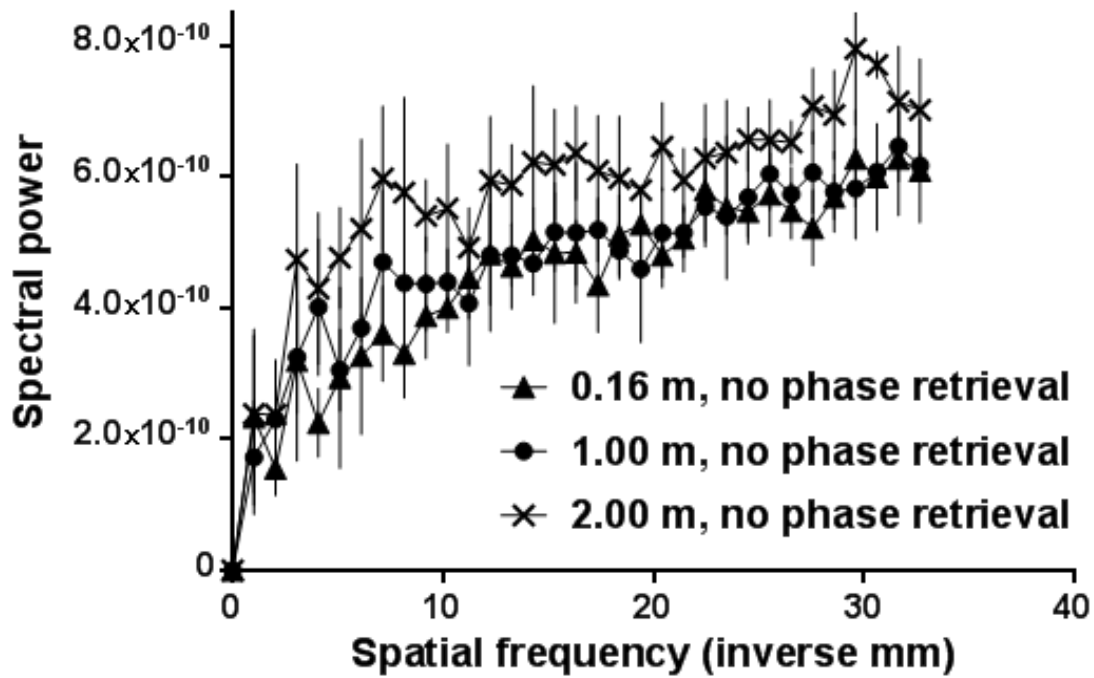

**Extended Data Figure 1: Power spectrum from CT showing image noise is essentially constant as a function of free space propagation distance**. Image noise is characterised by the spectral power at different spatial frequencies; here it is shown that additional free space propagation between the sample and detector does not increase image noise in the reconstructed CT (64 x 64 pixel ROIs). Uncertainties given by the standard deviation of repeated measurements on five CT slices.

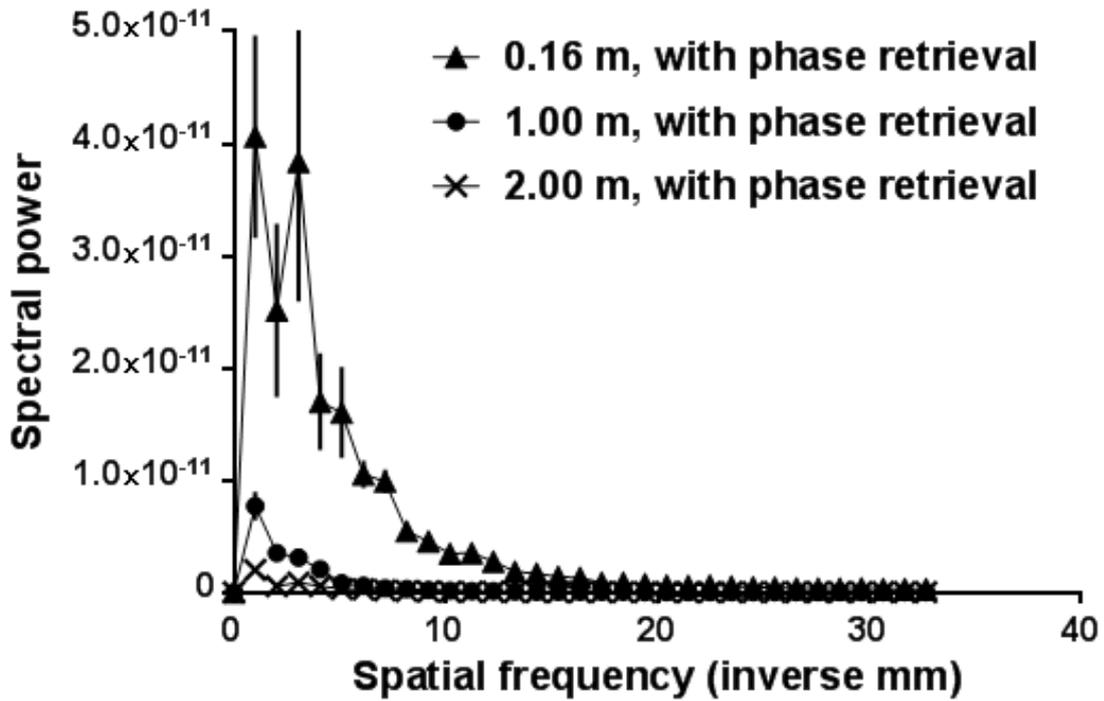

**Extended Data Figure 2: Power spectrum from CT showing noise suppression due to phase retrieval.** Here we see the effect of phase retrieval, causing a decrease in spectral power for the highest spatial frequencies and thus suppressing image noise (64 x 64 pixel ROIs). Uncertainties given by the standard deviation of repeated measurements on 5 CT slices.

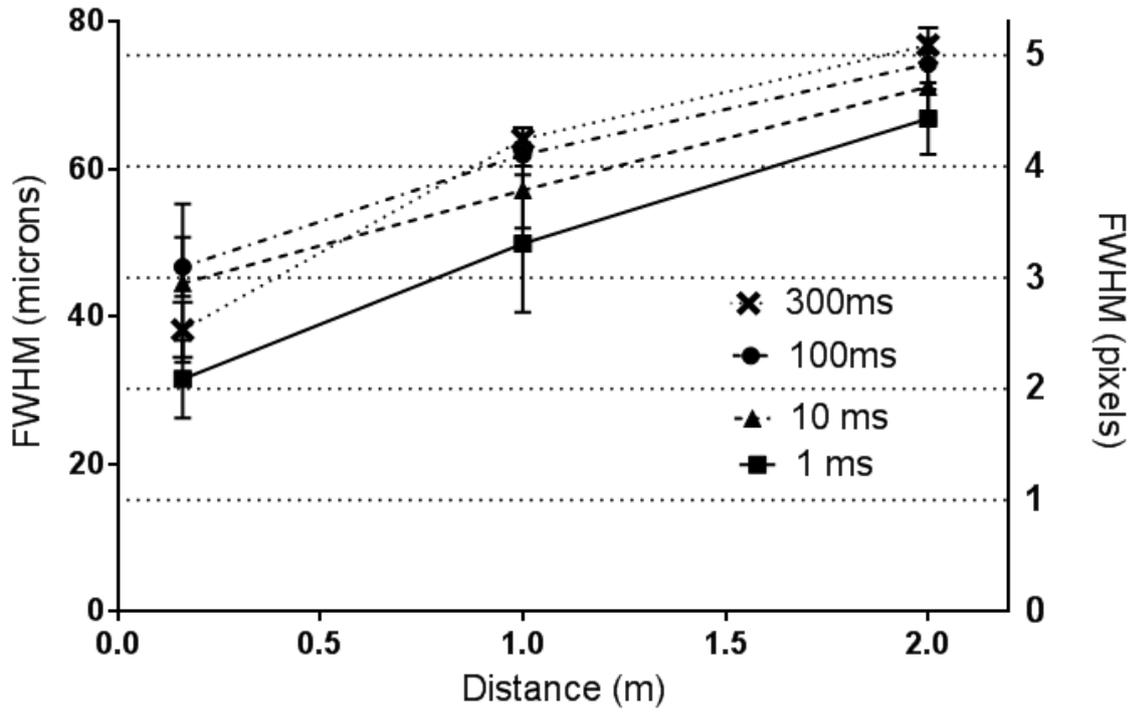

**Extended Data Figure 3: Spatial resolution of CT reconstructions.** Uncertainties represented by the standard deviation over n = 10 measurements. Exposure times were 1 ms (squares, solid line), 10 ms (triangles, dashed line), 100 ms (circles, dash-dotted line), and 300 ms (crosses, dotted line).